\documentclass[prd,preprint,showpacs,preprintnumbers,amsmath, nofootinbib]{revtex4}
\usepackage{color}
\usepackage[active]{srcltx}
\newcommand{\be}{\begin{equation}}
\newcommand{\ee}{\end{equation}}

\newcommand{\bse}{\begin{subequations}}
\newcommand{\ese}{\end{subequations}}
\newcommand{\bea}{\begin{eqnarray}}
\newcommand{\eea}{\end{eqnarray}}
\newcommand{\ba}{\begin{array}}
\newcommand{\ea}{\end{array}}

\def \th {\theta^{\mu\nu}}
\def\LLth{\theta^{-i}}

\def\nc{noncommutative\ }

\makeatother

\begin{document}
\preprint{HIP-2010-14/TH\cr IPM/P-2010/033}
\vspace*{3mm}
\title{ Light-Like Noncommutativity, Light-Front Quantization\\ and New
Light on UV/IR Mixing}
\author{{M. M. Sheikh-Jabbari$^1$ and A. Tureanu$^2$}}
\affiliation{$^1$School of Physics, Institute for research in
fundamental sciences (IPM), P.O.Box 19395-5531, Tehran, Iran}
\email{jabbari@theory.ipm.ac.ir}
\affiliation {$^2$Department of Physics, University of Helsinki and
Helsinki Institute of Physics, P.O.Box 64, FIN-00014 Helsinki,
Finland} \email{anca.tureanu@helsinki.fi}

\begin{abstract}

We revisit the problem of quantizing field theories on
noncommutative Moyal spacetime with \emph{light-like}
noncommutativity.  To tackle the issues arising from noncommuting
and hence nonlocal time, we argue that for this case light-front
quantization procedure should be employed. In this appropriate
quantization scheme we perform the non-planar loop analysis for the
light-like noncommutative field theories. One of the important and
peculiar features of light-front quantization is that the UV cutoff
of the light-cone Hamiltonian manifests itself as an IR cutoff for
the light-cone momentum, $p^+$. Due to this feature, the naive results of covariant quantization
for the light-like
case allude to the absence of the UV/IR mixing in the
light-front quantization. However, by a careful analysis of
non-planar loop integrals we show that this is not the case and the
UV/IR mixing persists. In addition, we argue in favour of the
perturbative unitarity of light-like noncommutative field theories in the light-front quantization scheme.

\end{abstract}
\pacs{11.30.Cp, 03.30.+p, 11.10.Nx}
\maketitle

\section{Introduction}

The notion of the usual (Lorentzian) space-time is expected to break
down due to quantum gravity effects. A simple way to model such
quantum effects in space-time, which is appearing in certain brane
and string theory settings \cite{SW}, is a space-time with
noncommuting
coordinate operators
\be\label{x-x-commutator}%
[x^\mu,x^\nu]=i\th ,%
\ee
where $\th$ is in general a function of $x^\mu$. In usual treatments,
which are again based on the picture coming from string theory
settings, $\th$ is a given matrix and is related to the vacuum
expectation value of a background Kalb-Ramond two form field
\cite{AAS}. As a result, the Lorentz symmetry on \nc (NC) space-time
is broken to one of its subgroups.

Lorentz invariance  has been at the heart of the formulation of quantum
field theories (QFTs), with their fundamental
properties such as causality, unitarity, spin-statistics and CPT
theorems, and the very notion of labeling the states (and the
fields) by representations of the Poincar\'e group. The lack of Lorentz
invariance could then jeopardize the pillars of QFT. Noncommutative
QFT with constant $\th$, though Lorentz-noninvariant, is symmetric
under the subgroup of Lorentz group which leaves invariant the
matrix $\th$, be it $O(1,1)\times SO(2)$ for space-like
noncommutativity ($\theta_{0i}=0$), $SO(1,1)\times SO(2)$ for
time-like noncommutativity ($\theta_{0i}\neq0$ ,
$\theta_{\mu\nu}\theta^{\mu\nu}<0$) \cite{LAG} or $T(2)$ for the light-like
case ($\theta_{\mu\nu}\theta^{\mu\nu}=0$). However, using the quantum-group concept of \emph{Abelian
twist} \cite{Drinfeld83} and relying on the translational invariance
of NC space-time with constant $\th$, it was shown that NC QFT has
a more subtle symmetry, under the \emph{twisted Poincar\' e algebra}
\cite{CKNT,CPrT,CNST}. This symmetry ensures that the particle
representations of NC QFT are identical with the representations of
the Poincar\'e algebra, thus preserving  the labeling of the quantum
states according to their mass and spin.

Equipped with the notion of twist, one can study QFTs  on NC
space-times \eqref{x-x-commutator} with constant $\th$ and still
rely on the CPT and spin-statistics theorems, which have been shown
to hold \cite{CPT,CNT,spin-stat}. The action for NC QFTs is obtained
from the commutative counterpart by replacing the usual product of
functions (fields) with the nonlocal Moyal $\star$-product:
\be\label{Moyal-product-def}%
(\phi\star\psi)(x)=\phi(x)\
e^{\frac{i}{2}\th\overleftarrow{\partial_\mu}\overrightarrow{\partial_\nu}}\
\psi(x)\ . %
\ee

The infinite nonlocality in the NC directions induced by the
$\star$-product can be an impediment for quantization. The
nonlocality in time is decisive: space-like NC field theories can be
quantized, while theories with NC time can not be quantized either
in interaction picture, or in Heisenberg picture or in path integral
formulation \cite{fujikawa}. This appears as a by-product of the
causality and unitarity analysis: it has been shown that time-like
NC field theories, upon naive covariant quantization, do not respect
(perturbative) unitarity \cite{GM}, nor causality \cite{SST}, while
space-like and light-like NC QFTs have been argued to be unitary
\cite{GM, AGM, LAG}.

Recently, light-like NC QFTs have been shown to be relevant in another context,
the Very Special Relativity \cite{VSR}. In \cite{VSR}, Cohen and Glashow put forward
the idea of a master theory at high energies, which includes the
effects of quantum gravity, preserving translational invariance but
breaking the Lorentz symmetry (and recovering it at lower energies), under the
name of \emph{Very Special Relativity}.
 They
 argued that one may relate the violation of the time-reversal, $T$,
and charge-conjugation times parity, $CP$, symmetries of particle
physics models with a possible violation of Lorentz invariance. This
can happen noting that there are subalgebras of Lorentz algebra
which, upon the action of $T$ and $CP$ on the generators, will close on
the whole Lorentz algebra (and not a subalgebra of it). One can show
that there are only four such possible subalgebras, the \emph{Very
Special Relativity} algebras, $T(2),\ E(2),\ HOM(2) $ and $SIM(2)$
\cite{VSR}. All of them are rank two and have only one-dimensional representations.

In \cite{NCVSR} we showed that NC and VSR invariant space-times are
related to each other:  $T(2)$ algebra and translational invariance
is the symmetry of a NC space-time with \emph{light-like
noncommutativity} ($\theta_{\mu\nu}\theta^{\mu\nu}=0$) and  constant
$\th$ (light-like Moyal plane);  the corresponding NC QFTs are then
formulations of $T(2)$ VSR quantum field theories. Actually, the
other three VSR subalgebras of Lorentz can also be realized as
symmetries of a light-like NC space-time, with certain $x$-dependent
noncommutativity parameter $\th(x)$, but with the violation of (a part of) translational
invariance, and thus physically not attractive.

In this new context, it is natural to investigate in more depth the light-like noncommutative cases, especially with regard to the
possibility of applying a consistent quantization approach.

It is interesting that, while being nonlocal in time, the light-like
case has turned out not to have a unitarity problem. The latter would ultimately
be related to the lack of a well-defined time-ordering needed for
computation of QFT correlation functions. This could presumably be traced to a
special feature in the light-like case, which other NC field theories
with nonlocal time do not share: in the light-cone frame, $\th$ has
only $\theta^{-i}$ components, where $i=1,2$ denotes the space
directions transverse to the light-cone,
 and hence the light-cone time coordinate $x^+$ is commuting with the other coordinates. In this sense, light-like noncommutative field theories in the light-front coordinates ({\it infinite-momentum frame}, \cite{FF,Weinberg,Susskind_LF}) are similar to the space-like noncommutative cases. In other words, the quantization of light-like noncommutative cases can be performed in the light-cone frame without a time-ordering ambiguity. For a review
on light-front quantization, see \cite{Brodsky-review}.

Having invoked the canonical quantization procedure for NCQFTs, one
can perform loop calculations. The effects of the NC phase resulting
from the $\star$-product \eqref{Moyal-product-def} in momentum space
does not show up in the tree level two point functions, but
the interaction terms are modified and hence, in general the NC effect appears in the
so-called non-planar loop Feynman diagrams \cite{MSV}. This
momentum-dependent phase factor improves the UV divergent behavior
of the non-planar loop diagrams, while giving rise to the
\emph{UV/IR mixing} phenomenon: sending the UV cutoff to infinity
leads to the appearance of undealt with IR divergences. The UV/IR mixing
seems to be a ubiquitous feature of all NCQFTs \cite{MSV, Susskind}
and persists beyond one loop \cite{A-Sh}.

In a nutshell, while the loop momentum of the planar diagrams is
regularized by a cutoff $\Lambda$, according to the standard result
of \cite{MSV}, in the non-planar loop diagrams we have a
corresponding effective cutoff $\Lambda_{eff}$ given by
\be\label{c-o} \Lambda_{eff}^2=\frac{1}{1/\Lambda^2+p\circ p}, \ee
where $p$ is the external momentum and $p\circ
p=(p^\mu\theta_{\mu\nu})^2$. The UV/IR mixing then arises from the
fact that upon taking the limit $\Lambda\to\infty$, the effective
cutoff becomes $\Lambda_{eff}^2=\frac1{p\circ p}$, which obviously
goes to infinity in the limit $p\to 0$ (IR limit). It should,
however, be noted that the expression \eqref{c-o} has been obtained
performing the loop integrals in the Wick rotated Euclidean frame.
This procedure, while being correct for space-like noncommutative
cases, as we have argued above and due to the subtleties with
time-ordering, will not be reliable for the light-like
noncommutative cases. Here, we show by explicit computations using
the light-cone time-ordering and light-front quantization, which is
the appropriate quantization scheme, that for the light-like cases
the dependence of $\Lambda_{eff}$ on external momenta is more
complicated than the one given in \eqref{c-o} and in particular the
dependence on the external momenta is not only a function of $p\circ
p$. In fact, if \eqref{c-o} were true for the light-like
noncommutative case we would have had good reason to believe that we
do \emph{not} have UV/IR mixing for the light-like case. As we shall
show in sections \ref{II} and \ref{III},  $p\circ p$ depends only on
the light-cone momentum $p^+$, which is regularized in the discrete
light-cone quantization (DLCQ) so that $p\circ p\sim \Lambda^{-2}$;
this could have led to the absence of UV/IR mixing.

Through a careful analysis of non-planar diagrams in the light-like
case in the DLCQ, which is the appropriate quantization scheme, we
show that the non-planar Feynmann diagrams do not remain IR finite
when the UV cutoff of the theory is removed and hence the UV/IR mixing
problem persists. Using our one loop results we argue for the
unitarity of the theory at this level.

\section{Discrete light-cone quantization and NC QFT}\label{II}

To study the
noncommutative non-planar loop diagrams
we consider the $\lambda\phi_\star^4$ theory with the action%
\be%
S=\int
d^4x\left[\frac12\partial_\mu\phi\partial^\mu\phi-\frac12m^2\phi^2-
\frac{\lambda}{4!}\phi\star\phi\star\phi\star\phi\right].%
\ee%
In the light-cone frame,%
\be%
ds^2=2dx^+dx^--dx^2_i,\quad \partial_\mu\phi\partial^\mu\phi=2\partial_+\phi\partial_-\phi-\partial_i\phi\partial_i\phi,
\qquad i=1,2,%
\ee%
where $x^\pm=\frac{1}{\sqrt2}(x^0\pm x^3)$, $x^+$ is taken to be the light-cone time and $x^i$ denote transverse directions. In the \emph{light-like NC case}, in this frame the only non-zero
component of $\th$ is $\theta^{-i}$ (or $\theta_{+i}$),
\be%
[x^-,x^i]=i\LLth.%
\ee%
In particular, we note that the $\theta^{+-}$ and $\theta^{+i}$ (hence
$\theta_{-+}$ and $\theta_{-i}$) components are zero. Consequently, the light-cone time coordinate commutes with all the other directions, $[x^+,x^-]=[x^+,x^i]=0$.  Note also that, due to the freedom to choose the $x_i$ axis in the transverse directions, there is only one physical observable related to $\theta^{-i}$, which can be taken to be $\theta^2\equiv \sum_{i=1,2} \theta^{-i}\theta^{-i}$.

In the \emph{space-like noncommutative case},
the light-cone frame can always be taken such that the
noncommutativity is only in the transverse directions, while in the
\emph{time-like case} the noncommutativity of light-cone time cannot
be avoided. Thus, for space-like and light-like noncommutative field theories there is a well-defined time-ordering
while for time-like we have time-ordering ambiguity.

As in the perturbative treatment of any QFT, we solve the equation
of
motion for the quadratic part of the action, whose solutions are%
\be%
\hspace*{-3.5mm}\phi(x)=\int\!\!\!
\frac{dp^+dp^2_\perp}{\sqrt{2(2\pi)^3p^+}}\left(\phi(p)
e^{i(p^+x^-+p^-x^+-p_ix_i)}+h.c.\right)%
\ee%
where $\vec{p}_\perp=(p^1,p^2)$  is the momentum transverse to the light-cone and%
\be\label{On-shell-p-}%
p^-=\frac{p^2_\perp+m^2}{2p^+},\quad p_\perp^2=p_1^2+p_2^2, \qquad p^+\geq 0\ . %
\ee%
$p^-$ should be viewed as the (eigenvalues for) the light-cone
Hamiltonian of the free theory and the non-negative $p^+$ is the
light-cone momentum, conjugated to $x^-$.
Taking the momentum conjugated to
$\phi$ to be $\frac{\partial {\cal L}}{\partial
(\partial_+\phi)}=\partial_-\phi$, we apply the canonical quantization procedure by imposing equal light-cone
time commutation relations between the field and its conjugate momentum. The commutator of the creation and annihilation operators for the scalar field is then obtained as
\be\label{LL-CCR}
[\phi(k^+,k_i),\phi^\dagger(p^+,p_i)]=\delta^2(k_i-p_i)\delta(k^+-p^+)\ .%
\ee

It was established long ago \cite{DLCQ} that light-front
quantization of ordinary  field theories leads to exactly the same
physical results as in any Lorentz frame. This is a
non-trivial result, if we recall that in the light-cone frame one cannot
use the standard Wick rotation to perform momentum integrals. The Feynman rules for light-front QFTs are slightly different than the
usual Lorentz covariant ones. These rules for general field theories
and gauge field theories in particular, have been worked out and
analyzed extensively  (see \cite{Brodsky-review} and references
therein).

In the light-front quantization there is a subtlety which should be
dealt with. As can be seen from \eqref{On-shell-p-}, in the
light-cone quantization there appear to be extra unphysical
``spurious'' poles at $p^+=0$, even at the tree level. To remove these
unphysical poles, one may employ the
 ``discrete light-cone quantization'' (DLCQ)
\cite{DLCQ, Brodsky-review}: by compactifying the $x^-$ direction on a circle
with radius $R_-$, the light-cone momentum $p^+$ takes quantized
values $n/R_-,\ n=0,1,2,\cdots$. One may then safely remove the now
isolated $p^+=0$ point and finally {in the end of all analysis}
remove $R_-$ by taking it to infinity. It is notable that $R_-$
which is an IR regulator for $p^+$ appears as a UV cutoff  for the
light-cone Hamiltonian (energy) $p^-$. This is an essential feature of the DLCQ, as will be seen later.

\section{One Loop Analysis in Light-Front quantization}\label{III}

In this section we turn on interactions given by  \cite{A-Sh}%
\be%
\begin{split}%
{S}_{int}=\frac{\lambda}{3\cdot 4!}\int & \prod_{n=1}^4\left(\frac{{d^4k_n}}{(2\pi)^4}\ \phi(k_n)\right)
(2\pi)^4\delta^4\left(\sum_{i=1}^4
 k_i\right)\times\\ & \times\biggl[\cos\frac{k_1\theta k_2}{2}
\cos\frac{k_3\theta k_4}{2}+\cos\frac{k_1\theta
k_3}{2}\cos\frac{k_2\theta k_4}{2}+\cos\frac{k_1\theta
k_4}{2}\cos\frac{k_2\theta k_3}{2}\biggr], %
\end{split}%
\ee%
where $k\theta p=\theta^{-i}(p^+k_i+k^+p_i)$. Consider the one loop
correction to the two-point function. Using standard light-front QFT
methods one can see that the loop integrals separate into two parts
\cite{MSV}: \emph{planar} parts which do not involve the NC
$\cos(k\theta p/2)$ factors and the \emph{non-planar} parts which do
involve $\theta$-dependent phases. The planar diagrams have the same
form as in ordinary field theory and can be regularized by a UV
cutoff $\Lambda$.

In what follows, in section \ref{space-like-section} we  discuss the planar and non-planar one loop correction to two point function for space-like  noncommutative case in the light-front quantization  and show that the result is the same as in the covariant quantization in which Wick rotation has also been used. Then, in section \ref{light-like-section} we address the same question for the light-like noncommutative case for which, as we have argued, the only consistent quantization scheme is the light-front quantization.

\subsection{Space-like NC case in light-front quantization}\label{space-like-section}

Let us start with  the \emph{planar two
point function}, for which the momentum integral\footnote{Since we are using DLCQ
prescription,  instead of the integral over $q^+$, we should have a
discrete sum. Hence, in our analysis, the $q^+$ integral is in fact an upper
bound
approximation to the loop expressions.} is \cite{A-Sh}:
\be%
I_P=\frac{2\lambda}{3}\cdot \frac{1}{2(2\pi)^3} \int^\Lambda d^2q\int_\epsilon^\Lambda \frac{dq^+}{2q^+}\,,%
\ee%
where $\Lambda$ is the UV cutoff and%
\be\label{loop-IR-cutoff}%
\epsilon=\frac{q^2_i+m^2}{\Lambda}\,,\ \ \ i=1,2.%
\ee%
Note that even for off-shell particles running in the loops, eq.
\eqref{On-shell-p-} is still valid and the integral is not taken
over $q^-$. Consequently, the UV cutoff on $q^-$ leads to an IR cutoff on $q^+$.
This integral can be readily  performed to obtain%
\be%
\begin{split}
I_P&=\frac{\lambda}{3}\cdot \frac{1}{2(2\pi)^3}\int^\Lambda d^2q\ \ln{\frac{q^2_i+m^2}{\Lambda^2}}\cr%
&=\frac{\lambda}{48 \pi^2}\left(\Lambda^2-m^2\ln\frac{\Lambda^2}{m^2}\right)+finite\
terms.
\end{split}
\ee%

For the \emph{nonplanar two-point function}, taking the
noncommutativity parameter in the
transverse space only, $\theta^{ij}\neq 0$ ($i,j=1,2$), the integral becomes:%
\be%
I^{SL}_{NP}=\frac{\lambda}{2\cdot 3}\cdot \frac{1}{2(2\pi)^3}
\int^\Lambda d^2q (e^{i\tilde p_i q_i}+e^{-i\tilde p_i q_i})
\int_\epsilon^\Lambda \frac{dq^+}{2q^+}\,,%
\ee%
where $\tilde p_i=\theta\epsilon_{ij}p_j$. We can perform the
integral by replacing the IR and UV cutoffs by exponentials, so that
the $q^+$
integral becomes%
\be%
I^+=\int^\infty_0\frac{dq^+}{q^+} e^{-\frac{q^+}{\Lambda}-\frac{\epsilon}{q^+}}=2K_0\left(\frac{2}{\Lambda}\sqrt{q^2+m^2}\right).%
\ee%
One then performs the angular part of the two-dimensional $q_i$
integral to obtain a Bessel function, yielding%
\be%
I^{SL}_{NP}=\frac{\lambda}{24\pi^2}\int_0^\Lambda\ dq\ q J_0(|\tilde
p|q)K_0\left(\frac{2}{\Lambda}\sqrt{q^2+m^2}\right)\,,%
\ee%
where $|\tilde p|^2=\theta^2 |p_\bot|^2$. Noting that for $q\gg \Lambda$
the $K_0$ function already involves an exponential damping and
that $J_0$ for large $q$ grows like $q^{-1/2}$, one may relax the UV
cutoff and use Formula 6.596-7 of \cite{G-R} to arrive at
\be%
\begin{split}%
I^{SL}_{NP} &=\frac\lambda{24\pi^2}\int_0^\infty\ dq\ q J_0(|\tilde
p|q)K_0\left(\frac{2}{\Lambda}\sqrt{q^2+m^2}\right)\cr &=
\frac{\lambda}{48\pi^2} m^2\cdot
\frac{\Lambda_{eff}}{m}K_{1}\left(\frac{2m}{\Lambda_{eff}}\right), %
\end{split}%
\ee%
where $ \Lambda_{eff}^{-2}=\Lambda^{-2}+\frac{|\tilde p|^2}{4}$.
This expression is exactly the same as what one would get using the
standard Wick rotation and doing Euclidean integrals \cite{MSV}.

\subsection{Light-like NC case in light-front quantization}\label{light-like-section}

As in the usual light-front quantization \cite{DLCQ} and as we have explicitly showed above, the planar diagrams are the same as the ones in the covariant quantization. We hence focus on the non-planar diagrams.
The
non-planar diagrams of the light-like case have to be computed anew
in the light-front quantization.
Using the standard light-front Feynman rules, the corresponding non-planar two-point function is%
\be\label{NP-full}%
{\cal
I}_{NP}^{LL}=I_{NP}^{LL}+(I_{NP}^{LL})_{c.c.},
\ee%
where
\be\label{NP-integral-LL-case}%
I_{NP}^{LL}=\frac{\lambda}{12\cdot (2\pi)^3}\int^\Lambda d^2q\
e^{i\alpha_i q_i} \int_\epsilon^\Lambda \frac{dq^+}{2q^+} e^{i\beta
q^+},
\ee%
\be\label{alpha-beta} %
\alpha_i=\theta^{-i}p^+, \qquad \beta=\theta^{-i}p_i, %
\ee%
and $(I_{NP}^{LL})_{c.c.}=I_{NP}^{LL}(-\alpha_i,-\beta)$.

 One of the
peculiar features of light-front field theory is that to remove the
pole at $p^+=0$, usually DLCQ procedure \cite{Brodsky-review} is
invoked, \emph{i.e.} we remove all the physical states with $p^+$
below $1/R_-$. Recalling \eqref{On-shell-p-}, this implies a lower
bound on the UV cutoff of the light-cone energy $m^2/(1/R_-)$. In
performing the loop integrals we impose the UV cutoff $\Lambda$ on
all momenta (a universal cutoff) and hence the IR tree-level cutoff
$R_-$ and the UV loop-level cutoffs are not independent and they
should be chosen such that
\be\label{cutoff-interplay}%
\Lambda \sim m^2R_-,  \quad \textrm{or} \quad
\frac{\Lambda}{R_-}\sim m^2.
\ee%
In other words,  the IR cutoff on the light-cone momentum running in the
loops is indeed the one in \eqref{loop-IR-cutoff}.

We perform the $q^+$ integral by replacing the IR and UV cutoffs by
exponentials, using the Formula 3.471-9 of \cite{G-R}:
\[
\begin{split} \int_\epsilon^\Lambda \frac{dq^+}{q^+} e^{i\beta
q^+}&=\int^\infty_0\frac{dq^+}{q^+}
e^{-\frac{q^+}{\Lambda}} e^{-\frac{\epsilon}{q^+}}\ e^{i\beta q^+}\\
&
=2K_0\left(\frac{2\sqrt{1-i\beta\Lambda}}{\Lambda}\sqrt{q^2+m^2}\right).%
\end{split}
\]
Computing the angular integral of the transverse $2D$ part
we obtain %
\bea\label{q-int}%
I_{NP}^{LL}&=&\frac{\lambda}{96\pi^2}\int_0^\infty\!\! q
K_0\left({2\sqrt{1-i\beta\Lambda}}\sqrt{\frac{q^2+m^2}{\Lambda^2}}\right)J_0(|\alpha|q)dq\nonumber
\\
&=&
\frac{\lambda}{2\cdot 96\pi^2} m^2\ \frac{\Lambda_{eff}}{m}K_{1}\left(\frac{2m}{\Lambda_{eff}}\right), %
\eea%
where %
\be%
\Lambda_{eff}^{-2}=\Lambda^{-2}+\frac{|\alpha|^2}{4}-\frac{i\beta}{\Lambda},
\quad |\alpha|^2=\theta^2(p^+)^2.%
\ee%
A proof of the above result which is based on the Formulae 6.596-7 and
8.535 of \cite{G-R} is given in the Appendix A.
The complete non-planar integral \eqref{NP-full} is then%
\be\label{24}%
{\cal I}_{NP}^{LL}=I_{NP}^{LL}(\Lambda_{eff})+I_{NP}^{LL}(\Lambda_{eff}^*).%
\ee
After exponentiating the IR and UV cutoffs on $q^+$, one can
perform the $q^i$ integral first and then the $q^+$ integral, and as
expected we get the same result as above.

When $|z|=\left|\frac{m}{\Lambda_{eff}}\right|$ is very small, one may approximate
$\frac{2}{z}K_1(2z)\simeq z^{-2}+\ln z^2$ and arrive at%
$$
I_{NP}^{LL}=\frac{\lambda}{2\cdot
96\pi^2}\left(\Lambda^2_{eff}-m^2\ln\frac{\Lambda_{eff}^2}{m^2}\right)\
$$
and, by using \eqref{24}, we obtain ${\cal
I}_{NP}^{LL}$ in terms of the real part and the absolute value of $\Lambda^2_{eff}$:
\be\label{Non-planar-LL-full}%
{\cal
I}_{NP}^{LL}=\frac{\lambda}{96\pi^2}\left[(\Lambda^2_{eff})_{Re}-m^2\ln\frac{|\Lambda_{eff}^2|}{m^2}\right]\
,
\ee%
where
\begin{eqnarray}\label{Lambda-eff-Re-norm}%
(\Lambda^2_{eff})_{Re}&=&\left(
{1+\frac{\Lambda^2|\alpha|^2}{4}}\right)
\left[\frac{1}{\Lambda^2}\left(1+\frac{\Lambda^2|\alpha|^2}{4}\right)^2+{\beta^2}
\right]^{-1},\cr%
|\Lambda^2_{eff}|&=&\Lambda\left[\frac{1}{\Lambda^2}\left(1+\frac{\Lambda^2|\alpha|^2}{4}
\right)^2+{\beta^2}\right]^{-1/2}.
\end{eqnarray}
Due to the appearance of the $\beta$ terms, the non-planar light-like
loop integral \eqref{Non-planar-LL-full}, unlike the non-planar
space-like noncommutative result \cite{MSV}, is not simply
a function of only $p\circ p=|\alpha|^2=\theta^{-i}\theta_{+i}(p^+)^2\equiv \theta^2 p^2$.%

\subsubsection*{UV/IR discussion}

To study  the UV/IR connection one should analyze
$|\Lambda^2_{eff}|$ and $(\Lambda^2_{eff})_{Re}$ in the
$\Lambda\to\infty$ limit, while the external momenta are held fixed
\cite{MSV}. In our case, however, as we discussed, we have the
peculiar feature that the minimum value of the external light-cone
momentum is indeed linked with the UV cutoff $\Lambda$. In
particular, \eqref{cutoff-interplay} implies that
\be\label{alpha-Lambda}%
{\Lambda}|\alpha| \gtrsim {m^2\theta}\ ,
\ee%
which must be taken into account when taking $\Lambda\to\infty$,
while $\beta=\theta^{-i}p_i$ is held fixed. Therefore, after sending
the UV cutoff $\Lambda$ to infinity one obtains:
\be\begin{split}\label{limit}%
(\Lambda^2_{eff})_{Re}\simeq &\ \left({1+\frac{m^4\theta^2}{4}}\right)\ \frac{1}{(\theta^{-i}p_i)^2}, \\
|\Lambda^2_{eff}|\simeq & \ \frac{\Lambda}{\theta^{-i}p_i}.\ %
\end{split}\ee%
Taking now the $p_i\to 0$ limit in  \eqref{limit}, we see that the expressions show IR divergences.

If the limits are taken in the opposite order, i.e. first the $p \to
0$ limit, we obtain, as in the case of space-space noncommutativity,
$\Lambda_{eff}=\Lambda$ \cite{MSV}. Again, the limits do not
commute, and the UV/IR mixing persists. To confirm the validity of our present result,
in the Appendix B we have performed the
integral \eqref{NP-integral-LL-case} using other regularization
methods.


\subsubsection*{Discussion on the unitarity of the light-like NCQFT}

Based on string theory results, it has been argued \cite{AGM} that
light-like NC field theories are unitary, in the sense of having a
unitary S-matrix. Time-like NC cases are, however, non-unitary as
they do not satisfy the Cutkosky rules (and the optical theorem)
\cite{GM}. A simple way to address the unitarity issue is to
consider  the one loop effective action focusing on the two point function
result, which in the standard case (with $\Lambda_{eff}$ given by \eqref{c-o}) is%
\be%
S_{1PI}=\int d^4p\
\frac12\left[p^2-M^2+\frac{\lambda}{96\pi^2}\frac{1}{p\circ p}
\right] \phi(p)\phi(-p)\ ,%
\ee%
where $M$ and $\phi(p)$ are respectively renormalized mass and
field. The unitarity of the theory at this one-loop level could be
physically traced to the existence (or absence) of ghosts: if
$p\circ p$ is positive (negative) the kinetic term will remain
positive (or can become negative). We next recall that $p\circ p$ is
positive for space-like or light-like noncommutative cases while is
negative for time-like noncommutative case.

As we have shown, in the light-like case one should invoke
light-front quantization and the expression for the one-loop
non-planar diagram changes. Nonetheless, from \eqref{limit} it is
evident that the first term in \eqref{Non-planar-LL-full} is
positive definite and hence one expects the light-like
noncommutative field theories in light-front quantization to be
unitary field theories.

\section{Discussion}

We have argued that the quantization of field theories with
light-like noncommutativity is consistent only in the light-front
quantization approach. The argument is based on the fact that the
light-cone time $x^+$ is commutative ($[x^+,x^-]=[x^+,x^i]=0,\ \
i=1,2$), therefore it is local and the canonical quantization
procedure can be directly applied. The space-space noncommutative
field theories can also be quantized using the light-front
formalism, while for time-space noncommutativity the light-cone time
is nonlocal, just like the physical time, and the light-front method
is expected to be no more consistent than the covariant
quantization.

We have performed the one-loop analysis and showed that in the case of
space-space noncommutativity, the light-front quantization and the
covariant quantization give, as expected, exactly the same
results. It is expected that this equivalence hold to all orders.

Prompted by the hint given by the covariant quantization
analysis and in particular that the effective cutoff of the theory
for non-planar loops is given by \eqref{c-o}, one could have hoped that the
UV/IR mixing were absent in the  light-like noncommutative case
in the discrete light-cone quantization approach. However, through a
rigorous light-front study of the one-loop two point function of the
$\lambda\phi_\star^4$, we have shown that this is not the case and
the non-planar diagrams exhibit IR divergences. Here, to illustrate
this point, we have presented a two point function analysis at one
loop. It is, however, easy to see that the non-planar one-loop four
point function is controlled by similar ``effective cutoffs''
\eqref{Lambda-eff-Re-norm}. We expect this one loop result to be
carried to all loops order. Moreover, it is expected that, as in the
space-like noncommutative case \cite{A-Sh}, the noncommutativity parameter
$\theta$ receive no loop corrections at higher order in perturbation.

We have also briefly discussed the unitarity of light-like
noncommutative field theories and argued that based on our one-loop
two-point function result, we expect these theories to be
(perturbatively) unitary and free of ghosts. This point deserves a
more thorough study.

In our discussions we have considered the four-dimensional
$\lambda\phi_*^4$ theory. The results and, in particular, the
persistence of UV/IR mixing are expected to be true  for any
light-like NC quantum field and gauge field theories.
Our calculations show that light-front quantization in the
light-like NC field theories leads to different expressions compared to the covariant quantization approach.
This may bring some new features and results
which are not seen in the covariant approach. Loop analysis should be revisited using the light-front
quantization, with special emphasis on the NC QED loop effects and the chiral
anomaly in chiral NC gauge field theories.

As shown in \cite{NCVSR}, light-like noncommutative field theories
are a framework for formulating VSR-invariant field theories. It is hence
interesting to study the implications of our results and analysis in
this context.

\section*{Acknowledgements}

We are grateful to Masud Chaichian for useful discussions and comments. We also thank Jos\'e Gracia-Bond\'ia and Peter Pre\v{s}najder for comments on the manuscript. The
support of the Academy of Finland under the projects no. 121720 and
127626 is acknowledged.

\appendix
\section{More details on the integral in Eq. (\ref{q-int})}

In computing the integral in eq. (20) Formula 6.596-7 of  \cite{G-R}
has been used. This formula, however, as it stands only applies to
real argument for Bessel-K and Bessel-J functions, while in eq.
(\ref{q-int}) we have complex valued arguments. Below we shall show
that for our case and using another formula of \cite{G-R} together
with 6.596-7 one can prove eq. (\ref{q-int}). To start with, let us
recall that the integral we want to take is of the form
\be%
\hspace*{-3mm} {\cal I} =\int_0^\infty\!\! q
K_0\left(|\gamma|e^{i\Phi}\sqrt{{q^2+m^2}}\right)J_0(|\alpha|q)dq,
\ee where \be \gamma^2\equiv
|\gamma|^2e^{2i\Phi}=\frac{4}{\Lambda^2}(1\pm i\beta\Lambda)\ . \ee
From the above, we learn that $-\frac{\pi}{2}\leq 2\Phi\leq \frac{\pi}{2}$. Next, recalling Formula 8.535 of \cite{G-R},%
\be\label{K-Bessel-expand}\begin{split}
K_0\left(|\gamma|e^{i\Phi}\sqrt{{q^2+m^2}}\right)  =&\sum_{n=0}^\infty \frac{1}{n!} K_n\left(|\gamma|\sqrt{{q^2+m^2}}\right) \times\\
&\times \left(\frac{1-e^{2i\Phi}}{2}|\gamma|\right)^n
({q^2+m^2})^{n/2},
\end{split}
\ee
the integral takes the form%
\be\label{integral-JK-n}\begin{split}
 {\cal I} &=\sum_{n=0}^\infty \frac{1}{n!} \left(\frac{1-e^{2i\Phi}}{2}|\gamma|\right)^n  \\
 & \times\int_0^\infty\!\! J_0(|\alpha|q)
K_{-n}\left(|\gamma|\sqrt{{q^2+m^2}}\right)({q^2+m^2})^{n/2} qdq\ ,
\end{split}\ee
where in the above we have used the fact that $K_{-n}(z)=K_n(z)$. We
should also comment that in using Formula 8.535 of \cite{G-R} we
need to make sure that $|1-e^{i\Phi}|^2<1$. This condition is
fulfilled if $\cos\Phi>\frac12$, which includes our case where
$-\frac{\pi}{4}\leq \Phi\leq \frac{\pi}{4}$.

The integral in \eqref{integral-JK-n} can now be safely taken using Formula 6.596-7 of  \cite{G-R} to obtain%
\be\label{The-sum}
 {\cal I}=\frac{m}{\sqrt{|\alpha|^2+|\gamma|^2}}\sum_{n=0}^\infty \frac{1}{n!} \left(\frac{1-\lambda^2}{2}\right)^n
 Z^n K_{n+1}(Z)
\ee where
\be\begin{split}%
Z &=m\sqrt{|\gamma|^2+|\alpha|^2}\ ,\\
\lambda^2&\equiv
1-(1-e^{2i\Phi})\frac{|\gamma|^2}{{|\alpha|^2+|\gamma|^2}}=\frac{|\alpha|^2+\gamma^2}{|\alpha|^2+|\gamma|^2}\
.
\end{split}\ee
In the above we have used the notation
$\gamma^2=|\gamma|^2e^{2i\Phi}$. \emph{Iff} $|1-\lambda|^2<1$, then
one can resum the above series using  Formula 8.535 of \cite{G-R} to
obtain \be\begin{split}
{\cal I}&=\frac{m}{\sqrt{|\alpha|^2+\gamma^2}}K_{1}(m\sqrt{|\alpha|^2+\gamma^2})\\ &=m^2\ \frac{\Lambda_{eff}}{2m}K_{1}\left(\frac{2m}{\Lambda_{eff}}\right),%
\end{split}\ee
where
\be%
\Lambda_{eff}^{-2}=\Lambda^{-2}+\frac{|\alpha|^2}{4}-\frac{i\beta}{\Lambda}.
\ee%
The only point one should check to complete the proof is whether
$|1-\lambda|^2<1$ holds. If $\beta\Lambda=x,\ |\alpha|\Lambda=2y$,
then $\lambda=|\lambda|e^{i\psi}$, where
$$
|\lambda|^2=\frac{\sqrt{(y^2+1)^2+x^2}}{y^2+\sqrt{x^2+1}},\
\tan2\psi=\frac{x}{y^2+1}\ .
$$
It is then easy to see that the condition $|\lambda|<2\cos\psi$
holds for the case of our interest in UV/IR discussion, namely $x$
large but $y$ finite (that is, in the $\Lambda\to\infty$ limit).

\section{Schwinger regularization method}

Here we use another regularization method which is based on the
Schwinger parametrization of the ``propagator'', that is by
\emph{exponentiating the propagator or the expression which gives
the poles.} In our case,
$$
\frac{1}{q^+}=\int_0^\infty dx\ e^{-xq^+}\ .
$$
The UV and IR cutoffs on $q^+$, respectively $\Lambda$ and
$\epsilon$, then turn to IR and UV cutoffs on $x$, namely $1/\Lambda$
and $1/\epsilon$. Let us now perform the $q^+$ integral:
\be\begin{split}%
I^+&=\int_\epsilon^\Lambda \frac{dq^+}{q^+} e^{-i\beta q^+}
=\int_{\frac{1}{\Lambda}}^{\frac{1}{\epsilon}} dx \int_0^\infty dq^+
e^{-xq^+}e^{-i\beta q^+}\\ &=
\int_{\frac{1}{\Lambda}}^{\frac{1}{\epsilon}}
dx\frac{dx}{x+i\beta}=-\ln{\frac{\epsilon}{\Lambda}}+\ln{\frac{1+i\beta\epsilon}{1+i\beta\Lambda}},
\end{split}\ee%
where $\beta$ is defined in \eqref{alpha-beta} and
$$\epsilon=\frac{q_i^2+m^2}{\Lambda}.$$

We can now perform the $q$-integrals. There are two such integrals:

\be\begin{split}
{\cal I}_1&=\int^\Lambda d^2q\ e^{i\alpha_i{q^i}}\ln{\frac{q^2+m^2}{\Lambda^2}}=\int^\Lambda d^2q\ e^{i\alpha_i{q^i}}\int_\epsilon^\Lambda \frac{dq^+}{q^+}  \\
&=2\pi m^2\cdot
\frac{\Lambda_{eff}}{m}K_{1}\left(\frac{2m}{\Lambda_{eff}}\right), %
\end{split}%
\ee%
where $$ \Lambda_{eff}^{-2}=\Lambda^{-2}+\frac{|\alpha|^2}{4}.$$
This part of the integral is $\beta$-independent.

Next, let us consider the $\beta$-dependent $q$-integral:
\be\begin{split} I_\beta &=\int^\Lambda d^2q\
e^{i\alpha_i{q^i}} \ln{\frac{1+\beta^2\epsilon^2}{1+\beta^2\Lambda^2}}\\
&= \pi \int^\Lambda qdq\ J_0(|\alpha|q) \ln{\frac{1+\beta^2\epsilon^2}{1+\beta^2\Lambda^2}},%
\end{split}\ee
which can be rewritten as \bse\begin{align} I_\beta &=
\pi\Lambda^2 \int_0^1 ydy\ J_0(ay) \ln{\frac{1+b^2\rho^2}{1+b^2}}\label{a}\\ %
&\simeq \pi\Lambda^2 \int_0^\infty ydy\ J_0(ay) e^{-y^2}\ln{\frac{1+b^2\rho^2}{1+b^2}},\label{b}%
\end{align}\ese
where \be\label{abrho} a=|\alpha|\Lambda,\quad b=\beta\Lambda,\quad
\rho=y^2+\frac{m^2}{\Lambda^2}. \ee We may use either of (\ref{a})
or (\ref{b}) to continue the analysis; here we focus on (\ref{a}).


In order to perform the $y$-integrals of $I_\beta$ it turns out to
be more convenient to integrate the expression before adding its
complex conjugate. We use the series expansion \be\begin{split}
\ln{\frac{1+ib\frac{\epsilon}{\Lambda}}{1+ib}} &=\ln\left(1-\frac{ib}{1+ib}\left(1-\frac{\epsilon}{\Lambda}\right)\right)\\
&=-\sum_{n=1}^\infty \frac{1}{n} \left(\frac{ib}{1+ib}\right)^n
\left(1-y^2-\frac{m^2}{\Lambda^2}\right)^n,
\end{split}\ee
where $b=\beta\Lambda$ and $y=q/\Lambda$. Then, \be I_\beta=\hat
I_\beta+\hat I_\beta^{c.c.}, \ee where
$$
\hat I_\beta=-\pi\Lambda^2\sum_{n=1}^\infty \frac{1}{n}
\left(\frac{ib}{1+ib}\right)^n\left(\sum_{k=0}^n
\left(\begin{array}{l}n\\ k\end{array}\right)
\left(-\frac{m^2}{\Lambda^2}\right)^kA_{n-k}(a)\right),
$$
with ({\emph{cf}. Formula 6.683.6 of \cite{G-R}) \be A_n(a)=\int_0^1
dy\ yJ_0(ay) (1-y^2)^n=\frac{2^n n!}{a^{n+1}} J_{n+1}(a). \ee We
obtain \be \hat I_\beta=-\pi\Lambda^2\sum_{n=1}^\infty (n-1)!
\left(\frac{2ib}{1+ib}\right)^n
\sum_{k=0}^n\frac{1}{k!}\left(-\frac{m^2}{2\Lambda^2}\right)^k\frac{J_{n-k+1}(a)}{a^{n-k+1}}
\ee and hence \be\begin{split}
I_\beta=-\pi\Lambda^2\sum_{n=1}^\infty &(n-1)! \left(\frac{2b}{1+b^2}\right)^n[(b+i)^n+(b-i)^n]\\
&\times
\sum_{k=0}^n\frac{1}{k!}\left(-\frac{m^2}{2\Lambda^2}\right)^k\frac{J_{n-k+1}(a)}{a^{n-k+1}}\
,
\end{split}\ee
or \be\begin{split} I_\beta=-2\pi\Lambda^2\sum_{n=1}^\infty &(n-1)!
\left(\frac{2b^2}{1+b^2}\right)^n\left[\sum_{j=0}^{[\frac n2]}
\frac{(-1)^j}{b^{2j}}
\left(\begin{array}{l}\ n\\ 2j\end{array}\right)\right]\\
&\times
\sum_{k=0}^n\frac{1}{k!}\left(-\frac{m^2}{2\Lambda^2}\right)^k\frac{J_{n-k+1}(a)}{a^{n-k+1}}.
\end{split}\ee

\subsection*{UV/IR analysis}

In order to detect the UV/IR mixing in the $\beta$-sector we study
the limit %
\be \Lambda\to\infty,\quad \beta=\mathrm{fixed}. %
\ee%
In this limit, $I_\beta$ decomposes into $\Lambda^2$ terms and
$\Lambda^0$ terms: \be\begin{split}
I_\beta &=-\pi\Lambda^2\sum_{n=1}^\infty (n-1)! 2^{n+1} \frac{J_{n+1}(a)}{a^{n+1}}\\
&+\pi\sum_{n=1}^\infty (n-1)!
2^n\left(m^2\frac{J_{n}(a)}{a^{n}}+\frac{n(n+1)}{\beta^2}\frac{J_{n+1}(a)}{a^{n+1}}\right)+\cdots
\end{split}\ee
The $\Lambda^2$ term, which should be subtracted off in e.g. MS
regularization, is simply $b$ independent. Note that $J_n(a)/a^n$
remains finite at small and large $a$ and the sum over $n$ is
convergent and finite. The $\Lambda^0$ term has $1/\beta^2$. This
$\beta$ dependence leads to the IR divergence in $p_i\to 0$ limit
and hence UV/IR mixing. This confirms the result we presented in section \ref{light-like-section}, in eqs. \eqref{Non-planar-LL-full} and
\eqref{Lambda-eff-Re-norm}.


\end{document}